\documentclass[aps,prl,amssymb,footinbib,superscriptaddress,twocolumn]{revtex4} 
\usepackage{graphicx}
\usepackage{amsmath}
\usepackage{color}
\usepackage{cancel}
\usepackage{ulem}

\newcommand{\be}{\begin{equation}}
\newcommand{\ee}{\end{equation}}

\begin{document}

\title{Emergence of Classical Rotation in Superfluid Bose-Einstein Condensates}

\author{Angela White}
\affiliation{Quantum Systems Unit, 
Okinawa Institute of Science and Technology Graduate University, Onna-son, Okinawa 904-0412, Japan.}

\author{Tara Hennessy}
\affiliation{Quantum Systems Unit, 
Okinawa Institute of Science and Technology Graduate University, Onna-son, Okinawa 904-0412, Japan.}

\author{Thomas Busch}
\affiliation{Quantum Systems Unit, 
Okinawa Institute of Science and Technology Graduate University, Onna-son, Okinawa 904-0412, Japan.}

\begin{abstract} {Phase transitions can modify quantum behaviour on mesoscopic scales and give access to new and unusual quantum dynamics. Here we investigate the superfluid properties of a rotating two-component Bose--Einstein condensate as a function of changes in the interaction energy and in particular through the phase transition from miscibility to immiscibility. We show that the breaking of one of the hallmarks of superfluid flow, namely the quantisation condition on circulation, is continuous throughout an azimuthal phase separation process and displays  intriguing density dynamics.  We find that the resulting currents are stable for long times and possess a phase boundary that exhibits classical solid body rotation, despite the quantum nature of superfluid flow. To support this co-existence of classical and quantum behaviour the system develops a unique velocity flow profile, which includes unusual radial flow in regions near the phase boundary. 
 } 
\end{abstract}
\maketitle

Phase transitions in quantum systems can have a dramatic impact on the quantum mechanical behaviour on mesoscopic scales. Superfluidity in Bose-condensed gases is a mesoscopic manifestation of quantum mechanical effects and one of its hallmarks is the existence of quantised flow around phase singularities as a response to external rotation \cite{Matthews1999,Madison2000,FetterVortRev2001,KevrekidisVortRev2004}. 
However, as the quantisation condition arises from the requirement of the single-valuedness of the wavefunction, an interesting, and less well investigated, generalization appears in superfluids composed of several components. In these systems, due to the interplay of intra- and inter-component interactions, the spinor order parameter can undergo a phase transition that modifies the global symmetry of the system. As the quantisation condition applies to each component independently, the path along which circulation is determined consequently depends on the presence of the other component. This has proven to be particularly striking in toroidally trapped  binary mixtures of BECs, where immiscibility can drive a transition to azimuthal phase separation, breaking the requirement of quantised circulation around the toroid \cite{Shimodaira2010}.  Below we show that this transition is continuous and leads to a phase boundary, which rotates as a classical solid body. While this might seem at first to be incompatible with the quantum nature of superfluid flow, this co-existence can be explained through the presence of a radial flow.
 
In superfluids the circulation around a closed path $\sf p$ is quantised according to  
$
\oint_{\sf p} \mathbf{v}\cdot \text{}d \mathbf{r} = n2\pi\hbar/m\,.
$
Here $n$ is an integer winding number, $m$ the atomic mass, $\hbar$ the reduced Planck constant, and the superfluid velocity field, $\mathbf{v}= \hbar \nabla \theta /m$, is completely determined by the gradient of the condensate phase, $\theta$. This implies the velocity field of a vortex has a tangential $1/r$ velocity profile,  in contrast to classical rigid-body rotation, where $\mathbf{v} = \mathbf{\Omega} \times \mathbf{r}$.
The creation of vortices is a response  to external rotation and depends, in particular, on the confining geometry. While in simply connected trapping potentials vortices with higher winding numbers are unstable \cite{highcharge}, multiply connected geometries are known to  support persistent currents with large angular momentum.

A simple multiply connected potential can be realised by a toroidal trap, which has recently been the subject of intense experimental interest \cite{Phillips2007,Eckel2014}. For single component condensates superflows have been shown to exist for up to $40$ s \cite{Campbell2011}, however  the superflow in toroidally trapped miscible two-component condensates has only been observed on much shorter timescales \cite{Beattie2013}.   

In this work we first study how the quantisation of circulation breaks down in the transition region between miscibility and phase separation in a rotating, toroidally trapped two-component condensate.  While deep in the phase separation regime angular momentum scales linearly with rotation frequency, close to the phase separation point an oscillatory behaviour is found, which is accompanied by significant changes in the order parameter.  At the same time, the phase profile, which drives the superfluid flow, adjusts in a way that allows quantum and classical behaviour to coexist.

The system is modelled using two coupled Gross--Pitaevskii (GP) equations, which aptly describe a two-component Bose--Einstein condensate in the limit of zero temperature. Each component is assumed to be a different hyperfine state ($j=1,2$) of the same atomic species,  $m_1=m_2=m$, \cite{Mertes2007,Beattie2013} 
and to consist of the same numbers of atoms, $N_1=N_2=N$.  
The two-dimensional coupled GP equations for the wave functions $\psi_{j}$ under rotation around the $z$-axis with rotation frequency $\vec{\Omega}=\Omega\vec{z}$, are then given by
\begin{equation} \label{gpecoupled}
i \hbar \frac{\partial \psi_{j}}{\partial t} =  \left(-\frac{\hbar^{2}}{2 m}\nabla^{2}+V_{j}+\sum_{i}^{1,2}Ng_{ji}|\psi_{i}|^2-\vec{\Omega} \cdot \hat{L}\right)\psi_{j} \,.
\end{equation}
In order to allow for stable systems with high angular momentum, we assume that the atoms are trapped in a harmonic ring-shaped potential of the form $V_{j}=\frac{1}{2}m\omega_{r}^{2}(r-r_{0})^{2}$, where $r_{0}$ is the toroidal radius, $r^2=x^2+y^2$, and $\omega_{r}$ is the radial trapping frequency, identical for both components.  Toroidal trapping potentials have recently been realised in several experiments by using, for example, a time-averaged harmonic potential with a Gaussian laser beam through its centre  \cite{Phillips2007} or all optical traps made by applying a  red-detuned  Laguerre-Gauss mode of a laser beam \cite{Campbell2011,Moulder2012,Beattie2013}.  Alternatively, ring-shaped trapping potentials which vanish asymptotically can be produced in the evanescent field of a nanofiber \cite{LeKien2004,Vetsch10}.

The coupling constants, $g_{ij} = \sqrt{8\pi}\hbar^{2}a_{ij}/(ma_{z})$, describe atom-atom interactions in terms of the three-dimensional scattering length $a_{ij}$, and the characteristic harmonic oscillator length in the $z$ direction, $a_{z}=\sqrt{\hbar/m\omega_{z}}$.  For simplicity we choose the s-wave scattering lengths within each component to be equal, that is $a_{11}=a_{22}=a$, and for both species to experience the same out-of-plane trapping frequencies, $\omega_{z}$.   The strength of atom-atom interactions  between the two components, $g_{12}$, will be varied to induce the phase transition. 

Homogeneous two-component condensates are miscible for values of $g_{12}^{2}< g_{11}g_{22}$ and immiscible or phase-separated when $g_{12}^{2}>g_{11}g_{22}$ \cite{PethickandSmith,Ao98,Timmermans98}.   For trapped condensates these values are slightly shifted due to the inhomogeneous density profile  \cite{Smyrnakis2009,Wen2012} and the density distribution in the phase separated regime is determined by the shape of the external trapping potential.
In narrow ring traps (and when $g_{11}=g_{22}$), azimuthal phase separation is favoured (see for instance \cite{Shimodaira2010,Abad2014,Mason2013}), while in wider toroidal traps 
concentric ring configurations can occur (see for instance \cite{Mason2013,Abad2014}).
In the following we  numerically solve the coupled GP equations by applying a pseudo-spectral second order Strang method with symmetric three-operator splitting  \cite{Samlan2014}.

\begin{figure}[tb]
\includegraphics[width=\columnwidth]{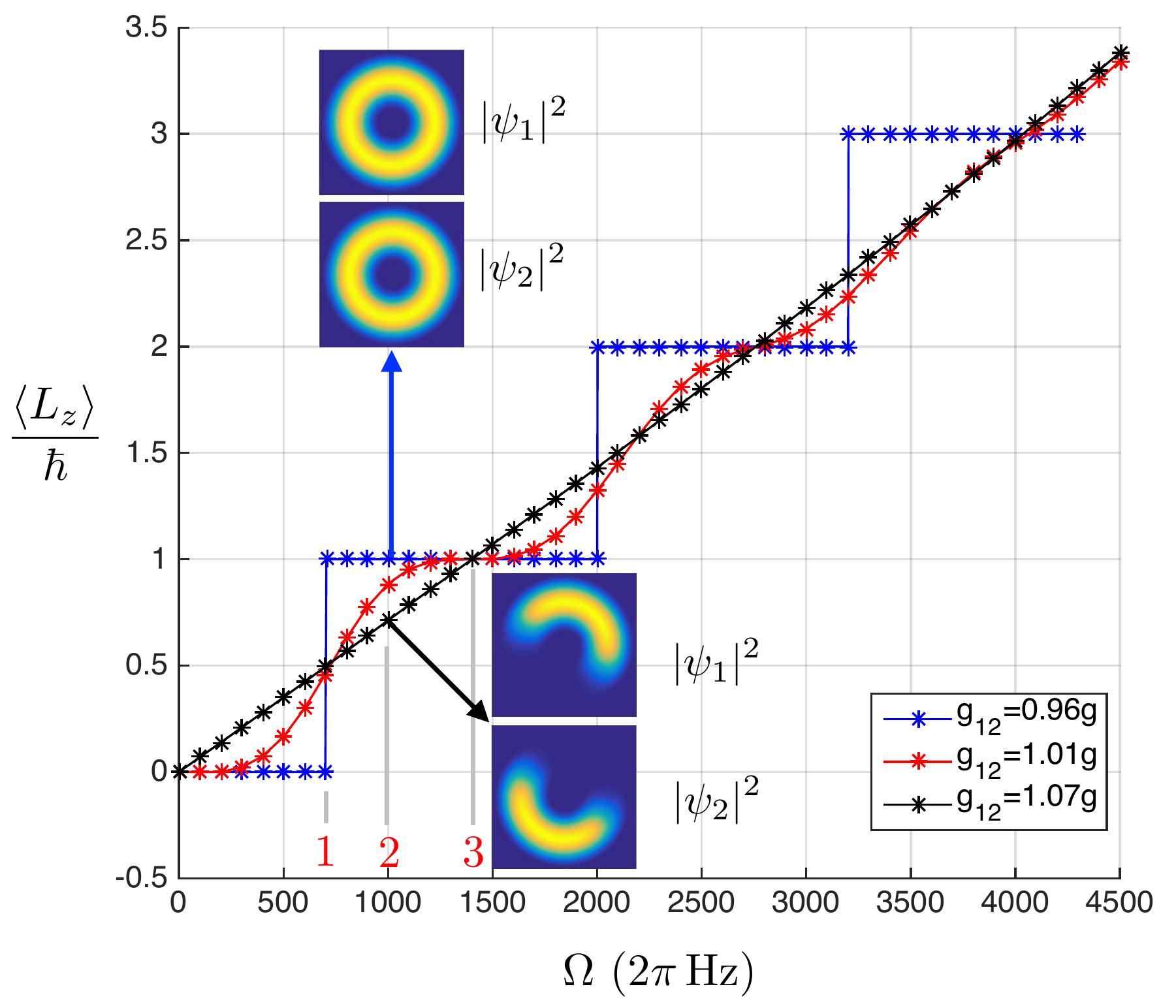}
\caption{Phase diagram mapping the transition from quantised steps of angular momentum to linear scaling of ${\langle L_{z}\rangle}/\hbar$ with $\Omega$, by varying $g_{12}$ from the miscible to immiscible regime. For $g_{12}$ on the border of miscibility to immiscibility, an oscillatory behaviour, damped with increasing $\Omega$ is observed. Simulation parameters: $\omega_{r}=2\pi \times 30000\,$Hz, $r_{0}=0.3\,\mu$m with $Ng = 1.1564\times10^{-41}\,$(Js$)^{2}$kg$^{-1}$. } \label{PhaseDiagram}
\end{figure}

\begin{figure}[tb]
\includegraphics[width=0.9\columnwidth]{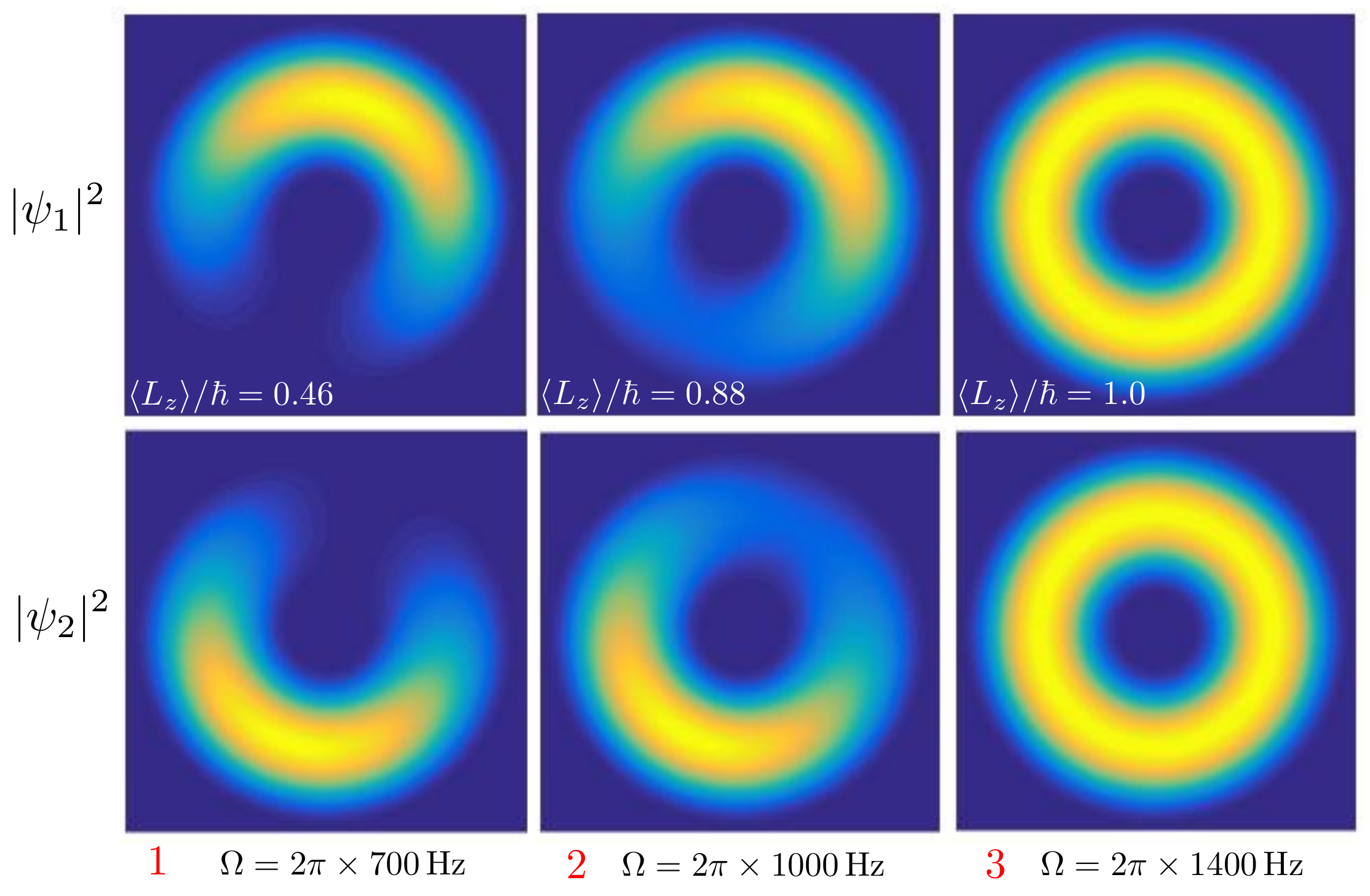}
\caption{Condensate density profiles in the transition between the miscible and immiscible regime. Simulation parameters: $g_{12}=1.01g$, $\omega_{r}=2\pi\times30000\,$Hz, $r_{0}=0.3\,\mu$m and $Ng = 1.1564\times10^{-41}\,$(Js$)^{2}$kg$^{-1}$. }
\label{fig:TransitionDensities}
\end{figure}

When the atom-atom interaction is chosen so that the two superfluid components are miscible, each component is multiply connected and circulation around the toroid is quantised. This implies that for each component the average angular momentum per particle,  $\langle L_{z} \rangle = i \hbar \int d \textbf{x} \, \psi_{j}^{*} \left(y\partial/\partial x-x \partial/\partial y \right) \psi_{j}$, is also quantised. 
In stark contrast, azimuthally phase-separated states break the multiply connected nature of each condensate component around the toroid and it was recently shown that they can therefore  rotate with arbitrary circulation and angular momentum of any value \cite{Shimodaira2010}. 

As experimentally realistic toroidal condensates are inherently of finite size, the phase-transition takes the form of a continuous cross-over and
in Fig.~\ref{PhaseDiagram} we show how the breakdown of the quantisation condition across this transition for 
the above system develops.  To do so, we calculate the angular momentum of the stationary state for three values of inter-atom interaction, $g_{12}$, selected such that the system is either fully miscible, fully immiscible or in the transition region between these two domains. As expected, for  $g_{12}$ well in the miscible regime,  quantisation of  angular momentum in each component  is observed (blue curve).  When $g_{12}$ is chosen so that the  condensate exhibits clear azimuthal phase-separation, angular momentum can be seen to scale linearly with $\Omega$ (black curve).  In the intermediate regime $\left(g_{12}\simeq g\right)$, however, an interesting damped oscillatory dependence of angular momentum with rotation frequency is found. The damping arises as a result of the increased rotation effectively cancelling the harmonic trapping potential, which shifts the value of the critical interspecies interaction strength at the phase separation point towards the free space result ($g_{12}^C=g$). For larger $\Omega$ the chosen interspecies interaction strength $g_{12}=1.01g$ therefore moves further into the phase separated regime and the curve becomes more linear. The density distributions corresponding to the miscible and immiscible regimes are also displayed in Fig.~\ref{PhaseDiagram}.

 On the border between miscibility and immiscibility (red curve in Fig.~\ref{PhaseDiagram}), the density distribution changes as a function of the rotation frequency and three examples corresponding to different angular momenta  are shown in Fig.~\ref{fig:TransitionDensities}.  This behaviour can be understood by realising that the rotational energy  acquired by the condensate is dependent on the frequency of the externally imposed rotation, $\Omega$. If the system is in a phase-mixed state at a rotation frequency that allows for an integer winding number, it can acquire a certain amount of rotational energy with increasing rotation frequency before it is energetically more favourable to phase separate and adjust the amount of angular momentum.  This leads to the observed cycling through mixed and phase separated  density distributions for condensates close to the phase boundary.

\begin{figure}[tb]
\includegraphics[width=\columnwidth]{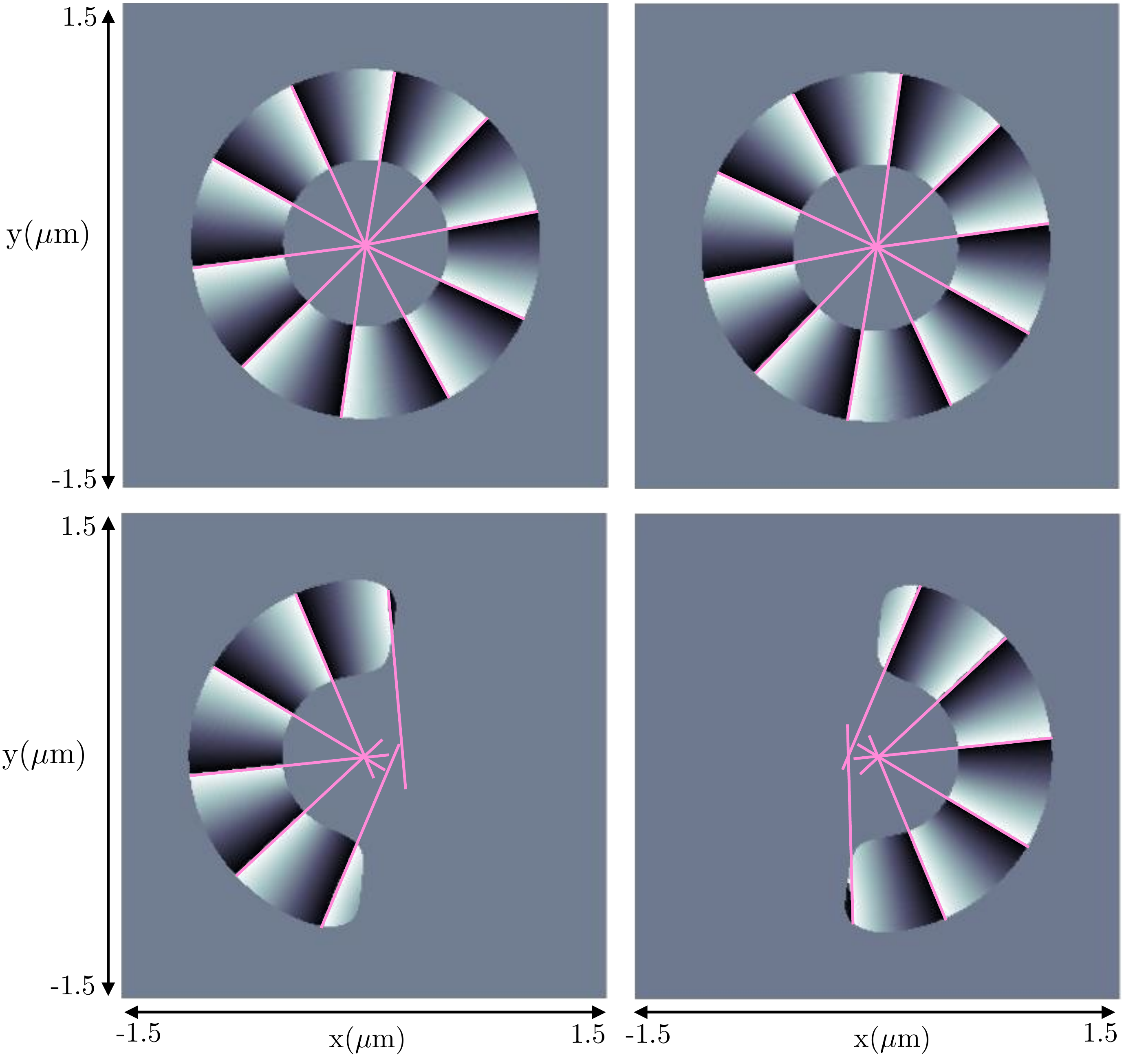}
\caption{Phase profile within the condensate, $\phi_{j}$, with overlaid lines of constant phase (pink). Upper row shows the two components in the miscible regime ($g_{12}=0.95g$), where the purely azimuthal flow is confirmed by the fact that all lines meet at a single  point. The lower row shows the phase separated regime ($g_{12}=1.6g$) and the presence of a radial flow component is indicated by the absence of a single crossing point.
Simulation parameters: $\Omega = 2\pi\times1910\,$Hz, $\omega_{r}=2\pi\times8000\,$Hz, $r_{0}=0.75\,\mu$m, $Ng = 1.1564\times10^{-41}\,$(Js$)^{2}$kg$^{-1}$.
 }\label{fig:PhaseSpokes}
\end{figure}

While in the mixing regime the well known $1/r$ velocity profile characteristic of rotation with quantised angular momentum is exhibited, it is easy to see that in the phase separated regime, where fractional winding numbers appear, this needs to be modified. In fact, if each superfluid component demonstrated a perfect vortex-like velocity profile everywhere, the phase boundaries would shear over time, as the atoms closer to the center of the potential move faster than those at the outer radial edges.  This would lead to an increase in the interaction energy and consequently unstable rotation.
Instead, to ensure that the phase boundary is always as short as possible, i.e.~along the radial direction, the system reacts by modifying the velocity profile away from purely azimuthal flow (see Fig.~\ref{fig:PhaseSpokes}). 

\begin{figure}[tb]
\includegraphics[width=1\columnwidth]{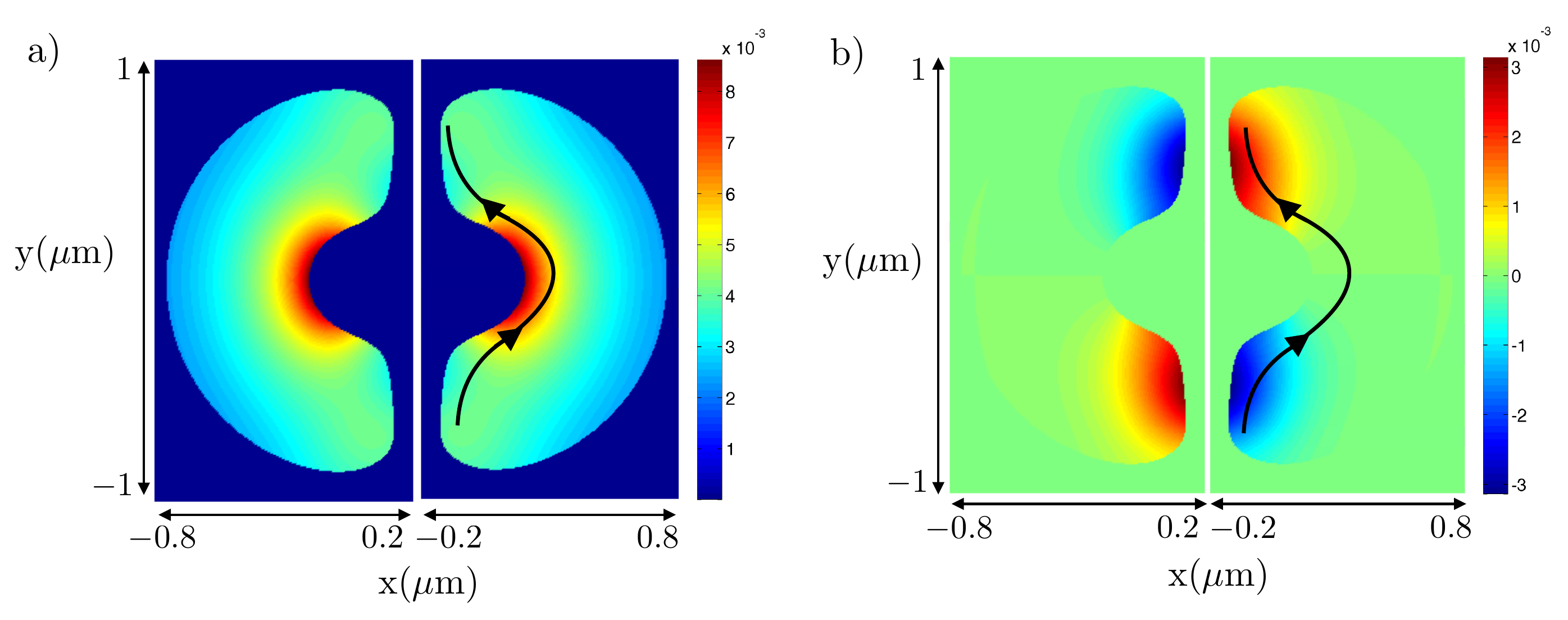}
\caption{Decomposition of the condensate velocity into (a) azimuthal and (b) radial components plotted within the condensate edge (defined as $6\%$ of the maximum density $|\psi_{j}|^{2}$). Simulation parameters: $\Omega = 2\pi\times1273\,$Hz, $\omega_{r}=2\pi\times10000\,$Hz, $r_{0}=0.5\,\mu$m, $Ng = 1.1564\times10^{-41}\,$(Js$)^{2}$kg$^{-1}$ and $g_{12}=1.2g$.}
\label{fig:veldecomp}
\end{figure}

To understand the flow profile in the phase separated case, we decompose the velocity field into its radial and azimuthal velocity contributions. These correspond to $v_{r}=\cos(\varphi)v_{x}+\sin(\varphi)v_{y}$ and $v_{\varphi} = -\sin(\varphi)v_{x}+\cos(\varphi)v_{y}$ and are shown in Fig.~\ref{fig:veldecomp}. Two regions where the velocity field exhibits distinctly unique behaviour can be clearly identified. In the bulk of each component, the flow displays the characteristic tangential superfluid velocity profile of the form $\mathbf{v}\propto n/r \,\mathbf{\hat{e}}_\varphi$, with $\mathbf{\hat{e}}_\varphi$ a unit vector in the direction of the azimuthal angle $\varphi$.  In contrast, in the vicinity of the phase boundary, the velocity field departs from a purely azimuthal profile and a radial flow develops. This is consistent with the fact that each component only reacts to the presence of the other over the scale of the healing length.

The appearance of the radial flow can be understood by realising that, in order to have minimal length, the phase boundary needs to move as a classically rotating object. This leads to a rotation velocity proportional to $r$, meaning that at larger radii the boundary has a larger velocity than at smaller radii. As this is in contrast to the $1/r$ velocity profile of the superfluid in the bulk region, the radial flow correctly re-distributes the atoms between the faster flow at smaller radii and the slower flow at larger radii. Behind the boundary, the radial flow leads to the movement of atoms from  smaller to larger radii, while at the same time reducing their azimuthal velocity, while in front of the phase boundary the opposite process takes place, with atoms flowing from larger to smaller radius (see Fig.~\ref{fig:veldecomp}).

The stable, classical solid body rotation of the phase boundary can be observed
when the external rotation is switched off and the system is able to evolve without constraint (see supplementary video 1).  As expected, the condensates rotate with an effective rotation frequency, $\Omega_\text{eff} = \langle v_{\varphi} / r \rangle$, which corresponds to the frequency of externally imposed rotation,  that is 
$\Omega/\Omega_\text{eff} \sim 1$. 

The above system is therefore an intriguing example where classical behaviour is displayed on a mesoscopic scale, despite the dynamics of the constituents being fully quantum mechanical. Similar behaviour can be found in the rotation of Abrikosov vortex lattices \cite{Abo-Shaeer20042001}. The emergence of large-scale classical behaviour in systems composed of quantum vortices also occurs in quantum turbulence, which  displays classical Kolmogorov scaling on length scales larger than the average inter-vortex spacing \cite{Maurer1998,Nore1997,Stalp1999,Araki2002,Salort2010PF}. 

In conclusion, we have studied the transition between miscibility and phase separation in rotating toroidally trapped two-component condensates. In the phase separated regime the requirement of quantisation of circulation is broken and azimuthally phase separated  superfluids can rotate with arbitrary circulation. However, to minimise the energy  of the system, the phase boundary has to always be aligned in the radial direction and therefore rotates as an effective solid body within the two-component flow. To resolve the dichotomy between this solid-body rotation, which has a velocity profile proportional to $r$, with the superfluid vortex profile in the bulk of the components, which is proportional to $1/r$, the system develops an unusual flow pattern involving radial components. This novel demonstration of the coexistence of classical and quantum behaviours can be observed in current state of the art cold-atom experiments. 

\begin{acknowledgements}
	We thank Marta Abad for valuable discussions. We acknowledge discussions with K.~Rza\.zewski, P.~Engels and L.~J.~O'Riordan. This work was supported by the Okinawa Institute of Science and Technology Graduate University. We are grateful to JSPS for partial support from Grant-in-Aid for Scientific Research (Grant No. 26400422).
\end{acknowledgements}


\end{document}